\documentclass[aps,pra,twocolumn]{revtex4-2}
\usepackage{graphicx}
\usepackage{color}
\usepackage{mathtools}
\usepackage{comment}
\usepackage{braket}
\usepackage{ulem}
\usepackage[subrefformat=parens]{subcaption}

\begin{document}

\title{Quantum Energy Teleportation and Entropy Change due to Feedback Control in One-Dimensional Heisenberg Model}
\author{Kanji Itoh$^{1}$, Yusuke Masaki$^{1}$, and Hiroaki Matsueda$^{1,2}$\thanks{hiroaki.matsueda.c8@tohoku.ac.jp}}
\affiliation{
$^{1}$Department of Applied Physics, Graduate School of Engineering, Tohoku University, Sendai 980-8579, Japan\\
$^{2}$Center for Science and Innovation in Spintronics, Tohoku University, Sendai 980-8577, Japan
}
\date{\today}

\begin{abstract}
We study the quantum energy teleportation in a four-spin one-dimensional Heisenberg model. A local magnetic field is applied at the edge sites to control the degree of the ground-state entanglement. In the teleportation protocol, an energy sender performs a projective measurement at one edge site, while an energy receiver performs a feedback control at the other edge site dependent on the measurement result to extract energy.  We find that the energy extracted by the receiver takes a maximum at intermediate value of the local magnetic field. We also find that this magnetic-field behavior is almost proportional to entropy changes due to the feedback control. The role of a feedback control in an entropy change is discussed in terms of entanglement thermodynamics.

\end{abstract}

\maketitle

\section{Introduction}
\label{sec:intro}
Quantum energy teleportation (QET) is a protocol for  transporting energy from one side to the other without using direct flow of quantum elementary excitations in solids~\cite{Hotta,Hotta2,Hotta3,Hotta4,Hotta5,Yusa,Hotta6,Frey,Trevison,Ikeda4}. 
Instead, the energy transportation is driven by nonlocal quantum entanglement between remote locations. 
Thus, one of the interesting physical issues behind the QET protocol is to clarify relationship between energy and quantum information, which is the central issue also in the field of information thermodynamics. 
In the presence of measurement and feedback, mutual transformation among information, work, and free energy is essential in the second law of information thermodynamics~\cite{text,Sagawa,Morikuni,Tajima}. 
The QET system provides a nice playground to examine the mechanism of the transformation. 
In addition to such an interest, a possibility of teleporting physical quantities opens a door for application of entanglement to quantum technologies such as utilization of large-scale quantum network due to long-range QET, energy allocation to a large number of remote nodes in quantum network, and purification of qubits by correlation-enhanced algorithmic cooling~\cite{Hotta6,Ikeda3,RB}.
Therefore, the QET protocol contains various aspects, ranging from fundamental physics to potential application in quantum devices. 

About 15 years ago, M. Hotta theoretically proposed, as the first QET protocol, the minimal QET model that consists of two qubits~\cite{Hotta,Hotta2,Hotta3,Hotta4,Hotta5}. He and his collaborators also examined QET in a quantum Hall system~\cite{Yusa}, a field-theoretical treatment of QET for understanding the long-distance behavior~\cite{Hotta6}, and temperature dependence of the QET performance in spin systems~\cite{Frey,Trevison}. Very recently, the QET has been experimentally demonstrated on a three-qubit system with nuclear magnetic resonance and on superconducting quantum hardware~\cite{2023,Ikeda1}. Furthermore, it has also been reported theoretically that the topological order in solids can be detected by the QET~\cite{Ikeda2}. Similar concepts have been investigated from the perspective of quantum information physics in recent years~\cite{Bruschi,Huber,Friis,Beny,Sapienza,Hackl}. For instance, it has been shown that there is finite energy cost when we get information about quantum entanglement in complex quantum systems~\cite{Beny}. The result may be viewed as a reverse process of QET. It is a good opportunity to examine the powerfulness and limitation of QET systematically. Here, we particularly focus on information thermodynamical aspects of QET.

The QET protocol is that an energy sender (Alice) and an energy receiver (Bob) perform local measurement and local feedback unitary operation, respectively, as well as their classical communication. Natural questions are how the maximum energy transfer is realized by optimizing the QET processes and also how the measurement and the feedback contribute to the optimization. These are the main purposes of this paper. 
We find that the maximum energy transfer is realized for a parameter characterizing the entanglement. 
The presence of the upper limit of the energy transfer reminds us with the second law of thermodynamics. 
The present case is formulated at zero temperature, but it is probably possible to establish an analogy to thermodynamics by utilizing the entanglement Hamiltonian in the ground state and the effective Hamiltonian after Bob's operation, which are defined from the Bob's reduced density matrix obtained by tracing out the Alice's degrees of freedom. We call this theory entanglement thermodynamics, and discuss the details in this paper.

The organization of this paper is as follows: in the next section, we formulate our QET protocol based on a quantum spin Hamiltonian and show the analytical results on the energies infused by Alice and obtained by Bob. 
We show numerical plots of the energies and various kinds of entropies in Sec.~\ref{sec:numericalresults} and discuss the origin of an entropy production in Sec.~\ref{sec:discussion}. Finally, we summarize our present work, and present future perspectives.

\section{Model and QET Protocol}
\label{sec:modelandprotocol}
Let us start with a four-spin quantum system with $S=1/2$.
The nearest-neighbor spins are coupled via antiferromagnetic Heisenberg interaction, and magnetic fields are applied at both edge sites. 
The Hamiltonian is defined by
\begin{eqnarray}
\label{H}
H=H_{A}+H_{B}+V, \label{eq:Htot}
\end{eqnarray}
where
\begin{align}
H_{A}&=hS_{1}^{z}+\vec{S}_{1}\cdot\vec{S}_{2}+\eta, \label{HA}\\
H_{B}&=hS_{4}^{z}+\vec{S}_{3}\cdot\vec{S}_{4}+\eta,\label{HB}\\
V&=\vec{S}_{2}\cdot\vec{S}_{3}+\xi\label{V}.
\end{align}
Here the parameters, $\eta$ and $\xi$, are determined so that
\begin{eqnarray}
{\rm tr} \left(\rho^{\rm{g}} H_A \right)={\rm tr} \left(\rho^{\rm{g}} H_B \right)={\rm tr} \left(\rho^{\rm{g}} V \right)=0,
\label{geta}
\end{eqnarray}
where $\rho^{\rm{g}}=\left|\psi\right>\left<\psi\right|$ is the density matrix of the ground state of $H$. 
Here and hereafter, the trace operation  ${\rm tr}(O)$ denotes the summation over the remaining degrees of freedom representing the operator $O$. As we introduce later, the subscript of tr accounts for the partial trace over the degrees of freedom specified by it. 
For $\alpha = x$, $y$, and $z$, the following commutation relations hold:
\begin{align}
    &\left[S_{1}^{\alpha},H_{B}\right]=\left[S_{1}^{\alpha},V\right]=0,\\
    &\left[S_{4}^{\alpha},H_{A}\right]=\left[S_{4}^{\alpha},V\right]=0.
\end{align}
In our QET protocol, we consider local operations on $\vec{S}_1$ and $\vec{S}_4$ by Alice (the sender) and Bob (the receiver), respectively. 
Therefore, 
the expectation value of $H_A$ ($H_B$) changes only by Alice's (Bob's) local operations, and that of $V$ never changes in the QET protocol. The latter fact means that no energy carrier propagates through the boundary of subsystems A and B. Here, we refer to two spins at site 1 and site 2 as subsystem A and two spins at site 3 and site 4 as subsystem B.

Our model contains two intermediate spins in addition to two edge spins which are controlled by Alice and Bob. Later we will introduce entanglement-oriented quantities in which we must decompose the total system into two subsystems for their definition. Then, we have some possibilities for selecting the position of the bipartition. This is the main difference from the minimal QET model in which the method of the bipartition is uniquely determined. Then, we can examine the role of the nearest neighbor spin of the Bob's control spin in the QET performance. We will mainly consider the central bipartition and the subsystem A is traced out. In this case, the remaining subsystem B contains two different types of spins. One is the edge spin at site 4 that is directly controlled by Bob, and the other at site 3 couples with the edge spin but Bob does not operate it. The Bob's local energy extracted from the system contains two components: the magnetic field term, and  the interaction term between spins 3 and 4. To distinguish them is a key for understanding the mechanism of the QET in our model. We can also consider the bipartition where only the site 4 is separated from the other sites. 
As shown later, the result of this bipartition is different from that of the central bipartition. On the basis of the difference, we clarify what kind of interaction dominates the transfer energy in the QET protocol.

The ground state of the Hamiltonian is represented as
\begin{align}
\label{psi}
\left|\psi\right>=&a\left(\left|\uparrow\downarrow\uparrow\downarrow\right>+\left|\downarrow\uparrow\downarrow\uparrow\right>\right)
-b\left(\left|\uparrow\uparrow\downarrow\downarrow\right>+\left|\downarrow\downarrow\uparrow\uparrow\right>\right) \nonumber \\
&-c\left|\uparrow\downarrow\downarrow\uparrow\right>-d\left|\downarrow\uparrow\uparrow\downarrow\right> ,
\end{align}
where the positive coefficients, $a$, $b$, $c$, and $d$ are determined by the diagonalization of the Hamiltonian matrix and their $h$ dependence is obtained. Using these coefficients, $\eta$ and $\xi$ are written as 
\begin{align}
    \eta&=\frac{1}{4}\left\{1-4b^2-2h(c^2-d^2)+4a(c+d)\right\},\\
    \xi&=\frac{1}{4}(2a^2+2b^2-c^2-d^2).
\end{align}
In this paper, we assume $0\le h < 1$ where $S_{\mathrm{tot}}^{z}=0$. 
Because the $z$-component of the total spin, $\sum_i S_i^z$, commutes with the total Hamiltonian \eqref{eq:Htot}, its eigenvalue denoted by $S_{\mathrm{tot}}^{z}$ is introduced as a quantum number. For $0 \le h < 1$, the ground state $\ket{\psi}$ is an entangled state and thus satisfies the necessary condition for QET. For $h>1$, the ground state is characterized by $S_{\mathrm{tot}}^{z}=-1$. In this case also, there is finite entanglement in the ground state $\left|\psi\right>$, but the entanglement mainly originates in the tendency of spin singlet formation between the intermediate spins $2$ and $3$ due to the interaction term $V$. The degree of entanglement can be controlled by the local magnetic field $h$. The entanglement between A and B is measured by the entanglement entropy $S(\rho_{34}^{\rm{g}})$, where $\rho_{34}^{\rm{g}}={\rm tr}_{12}\rho^{\rm{g}}$ and we define $S(\rho)=-{\rm tr}\left(\rho\log{\rho}\right)$.

The local operations performed by Alice and Bob are a projective measurement of spin 1 and a local unitary transformation of spin 4, respectively. The measurement operator with its output $\mu = 0$, $1$ is given by 
\begin{align}
\label{PA}
P_{1}(\mu)&=\frac{1}{2}\left\{ 1+(-1)^{\mu} \sigma^{x}_{1} \right\},
\end{align}
where $\sum_{\mu=0, 1}P_{1}(\mu)=1$ and we have introduced the  Pauli matrices $\sigma_i^{\alpha} = 2 S_i^{\alpha}$. 
The Bob's unitary operator is given by 
\begin{eqnarray}
\label{UB}
U_{4}(\mu)=\cos\theta+i (-1)^\mu\sigma^y_{4}\sin\theta,
\end{eqnarray}
where $\theta$ represents a rotation angle of spin at site 4 around the $y$-axis. We determine $\theta$ so that the extracted energy is maximized.
The measurement-dependent Bob's operation, $U_{4}(\mu)$, is feedback control. In Appendix~\ref{sec:appA}, we treat the case where Alice and Bob perform general projective measurements and general feedback unitary operations, respectively.

The typical protocol of QET is composed of the following three steps: $\rm(\hspace{.18em}i\hspace{.18em})$ Alice performs the projective 
 measurement $P_{1}(\mu)$ to the spin 1, and obtains the measurement result $\mu$. Through this measurement, the positive energy $E_{A}$ is infused to the subsystem $A$ on average. $\rm(\hspace{.08em}ii\hspace{.08em})$ The result $\mu$ is announced to Bob from Alice through classical communication. $\rm(i\hspace{-.08em}i\hspace{-.08em}i)$ Finally, Bob performs the unitary operation dependent on $\mu$, $U_{4}(\mu)$, and obtains positive energy $E_{B}$. 

Let us calculate the infused and the extracted energies in the above protocol. 
The density matrix of the state after step $\rm(\hspace{.18em}i\hspace{.18em})$ with measurement result $\mu$ is given by
\begin{eqnarray}
\rho^{\rm{m}}(\mu)=\frac{1}{p_\mu}P_1(\mu)\rho^{\rm{g}}P_1(\mu),
\end{eqnarray}
where $p_\mu={\rm tr}\left(\rho^{\rm{g}}P_1(\mu)\right)$ is the probability of obtaining result $\mu$. Since the energy of the system changes from ${\rm tr}\left(\rho^{\rm{g}}H\right)=0$ to $\sum_{\mu}p_\mu{\rm tr}\left(\rho^{\rm{m}}(\mu)H\right)$ on average, the infused energy $E_A$ is computed as
\begin{align}
    E_A&=\sum_{\mu}p_\mu{\rm tr}\left(\rho^{\rm{m}}(\mu)H\right)\nonumber\\
    &=\frac{1}{4}\left\{1-4b^2-2h(c^2-d^2)+2a(c+d)\right\},
\label{EA1}
\end{align}
where we have used $\mathrm{tr}\left(\rho^{\mathrm{m}}(\mu)H \right)=\mathrm{tr}\left(\rho^{\mathrm{m}}(\mu)H_A\right)$ and thus $E_{A}$ is independent of $H_{B}$ and $V$. Alice's measurement does not only infuse the energy but also reduces entanglement entropy of the subsystem B by
\begin{equation}
I_{QC}=S(\rho_{34}^{\rm{g}})-\sum_{\mu}p_\mu S(\rho_{34}^{\rm{m}}(\mu)),
\end{equation}
where $\rho_{34}^{\rm{m}}(\mu)={\rm tr}_{12}\rho^{\rm{m}}(\mu)$. 
The quantity $I_{QC}$ describes the information about the subsystem B that has been obtained by Alice's measurement and is called QC-mutual information (or Groenewold-Ozawa information) in information thermodynamics~\cite{Sagawa,Groenewold,Ozawa}. Note that Bob cannot obtain any information about the subsystem B unless Alice communicates the measurement results $\mu$ to him, i.e, $\sum_{\mu}p_\mu \rho_{34}^{\rm{m}}(\mu)=\rho_{34}^{\rm{g}}$ and thus, $S(\sum_{\mu}p_{\mu}\rho_{34}^m(\mu))=S(\rho_{34}^{\rm{g}})$.

Bob's operation dependent on $\mu$ transforms the density matrix as
\begin{eqnarray}
\rho^{\rm{f}}=\sum_{\mu}U_{4}(\mu)P_{1}(\mu) \rho^{\rm{g}} P_{1}(\mu)U_{4}^{\dagger}(\mu).
\end{eqnarray}
The total energy for this state is given by $\mathrm {tr}\left(\rho^{\mathrm{f}} H\right)=E_{A}+\mathrm{tr}\left(\rho^{\mathrm{f}} H_{B}\right)$. The second term accounts for the reduction of energy by the Bob's operation because $\mathrm{tr}(\rho^{\mathrm{g}}H_B) = 0$, and thus the extracted energy is given by
\begin{align}
E_{B}&=-{\rm tr}\left(\rho^{\rm{f}} H_{B}\right) \nonumber \\
&=-\sum_{\mu}\left<\psi\right|P_{1}(\mu)U_{4}^{\dagger}(\mu)H_{B}U_{4}(\mu)P_{1}(\mu)\left|\psi\right> \nonumber \\
&=\frac{1}{16}\sum_{\mu}\left[ X\cos\left(2\theta\right)+Y\sin\left(2\theta\right)-X \right] \nonumber \\
&\le\frac{1}{8}\left[\sqrt{X^{2}+Y^{2}}-X\right] = E_{B}^{\max},
\end{align}
where $X$ and $Y$ are given by
\begin{align}
X&=2-8b^2+4a(c+d)-4h(c^2-d^2),\nonumber\\
Y&=4b(4ha+c-d),
\end{align}
and the equality holds at $\theta=\frac{1}{2}\arctan{\frac{Y}{X}}$.
Note that $E_B$ can be positive only if Bob's unitary operation $U_4(\mu)$ depends on the Alice's measurements result.
Bob's feedback control changes the von Neumann entropy of the subsystem B by 
\begin{equation}
\Delta S_{34}=S(\rho_{34}^{\rm{f}})-S(\rho_{34}^{\rm{g}}), \label{eq:entropychange}
\end{equation}
with $\rho_{34}^{\rm{f}}={\rm tr}_{12}\rho^{\rm{f}}$.
From the convexity of $S(\rho)$ and the relation $S(U_4(\mu)\rho_{34}^{\rm{m}}(\mu)U_4^\dagger(\mu))=S(\rho_{34}^{\rm{m}}(\mu))$, we obtain an inequality
\begin{eqnarray}
    -\Delta S_{34}\le I_{QC},
    \label{second}
\end{eqnarray}
which
leads to the second law of information thermodynamics~\cite{Sagawa}.

\section{Numerical Results}
\label{sec:numericalresults}
In this section, we present numerical results for infused energy $E_A$ and maximally extracted energy $E_B^{\max}$ as well as the entanglement entropy 
 in the ground state $S(\rho_{34}^\mathrm{g})$ and the QC-mutual information $I_{QC}$. We then numerically clarify relationship between $E_B^{\max}$ and $\Delta S_{34}$.

\begin{figure}[tb]
\begin{center}
\includegraphics[width=8.5cm]{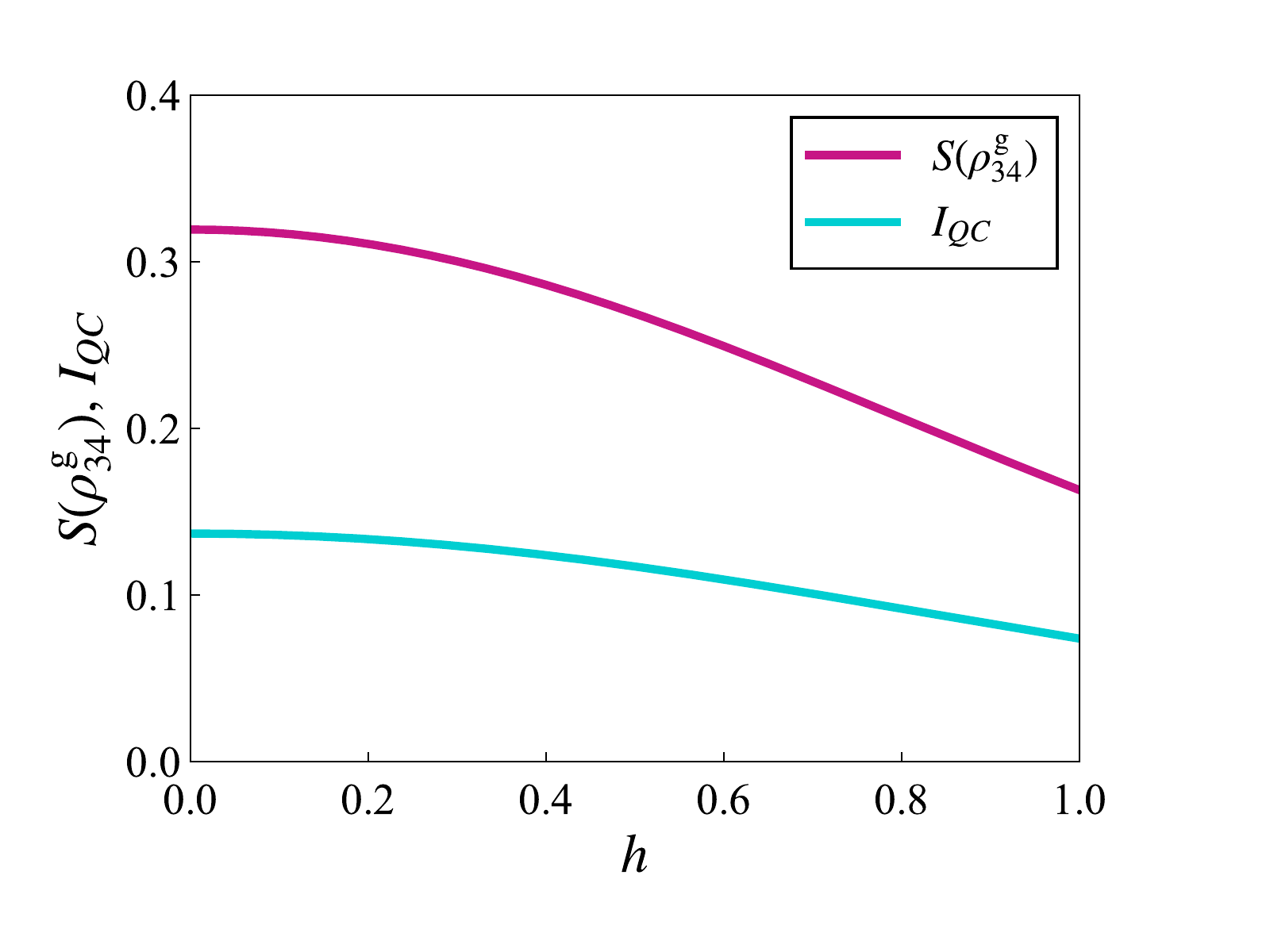}
\caption{
$S(\rho_{34}^{\rm{g}})$ and $I_{QC}$ as a function of magnetic field $h$.
}
\label{S34}
\end{center}
\end{figure}

Figure~\ref{S34} shows $S(\rho_{34}^{\rm{g}})$ and $I_{QC}$ as a function of the magnetic field $h$. 
We find that both of $S(\rho_{34}^{\rm{g}})$ and $I_{QC}$ monotonically decrease with $h$. 
The quantity $I_{QC}$ is always positive, and is bounded by the relation $I_{QC}<S(\rho_{34}^{\rm g})$.

\begin{figure}[t]
\begin{center}
\includegraphics[width=8.5cm]{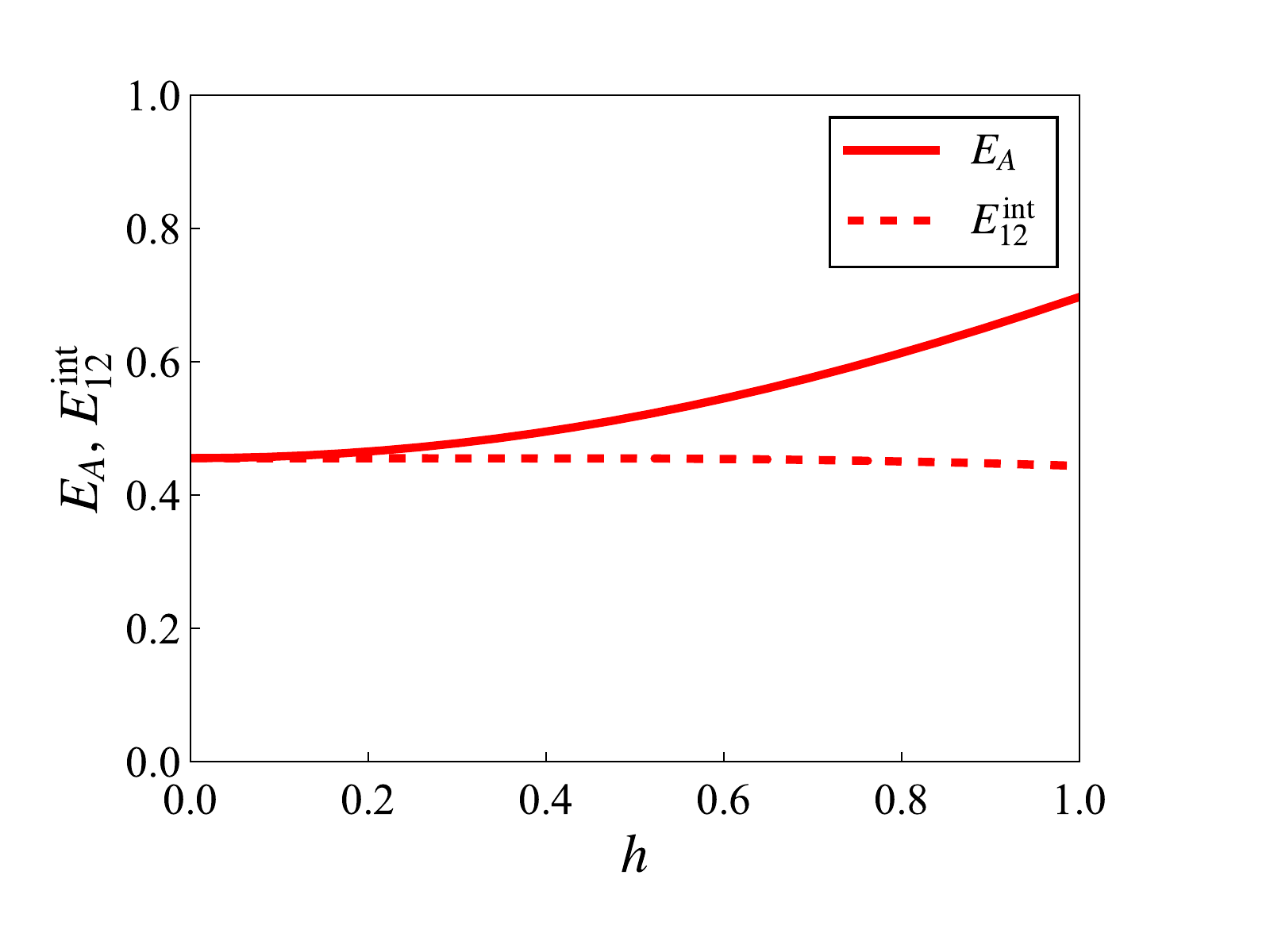}
\vspace{-0.8cm}
\caption{
$E_{A}$ and $E_{12}^{\rm{int}}$ as a function of magnetic field $h$. 
}
\label{EA}
\end{center}
\end{figure}

Next, we present the infused energy $E_A$ as a function of $h$ in Fig.~\ref{EA}. We find that  $E_A$ is positive and increases with $h$.
The infused energy $E_A$ has two different contributions according to Eq.~\eqref{HA}: the magnetic-field term and the spin interaction term. 
The latter is represented by
\begin{eqnarray}
E_{12}^{\mathrm{int}} = \sum_{\mu}p_{\mu}{\rm tr}(\rho^{\rm m}(\mu)\vec{S}_{1}\cdot\vec{S}_{2}) 
- \mathrm{tr}(\rho^{\mathrm{g}}\vec{S}_{1}\cdot\vec{S}_{2}),
\end{eqnarray}
and also shown in Fig.~\ref{EA}. The interaction energy $E_{12}^{\mathrm{int}}$ infused via the measurement is almost independent of $h$, and thus the magnetic-field term is essential in the $h$-dependence of $E_A$. Note that $E_{A}$ is not linearly dependent on $h$, although $h$ is a coefficient of the magnetic-field term.

\begin{figure}[t]
\begin{center}
\includegraphics[width=8.5cm]{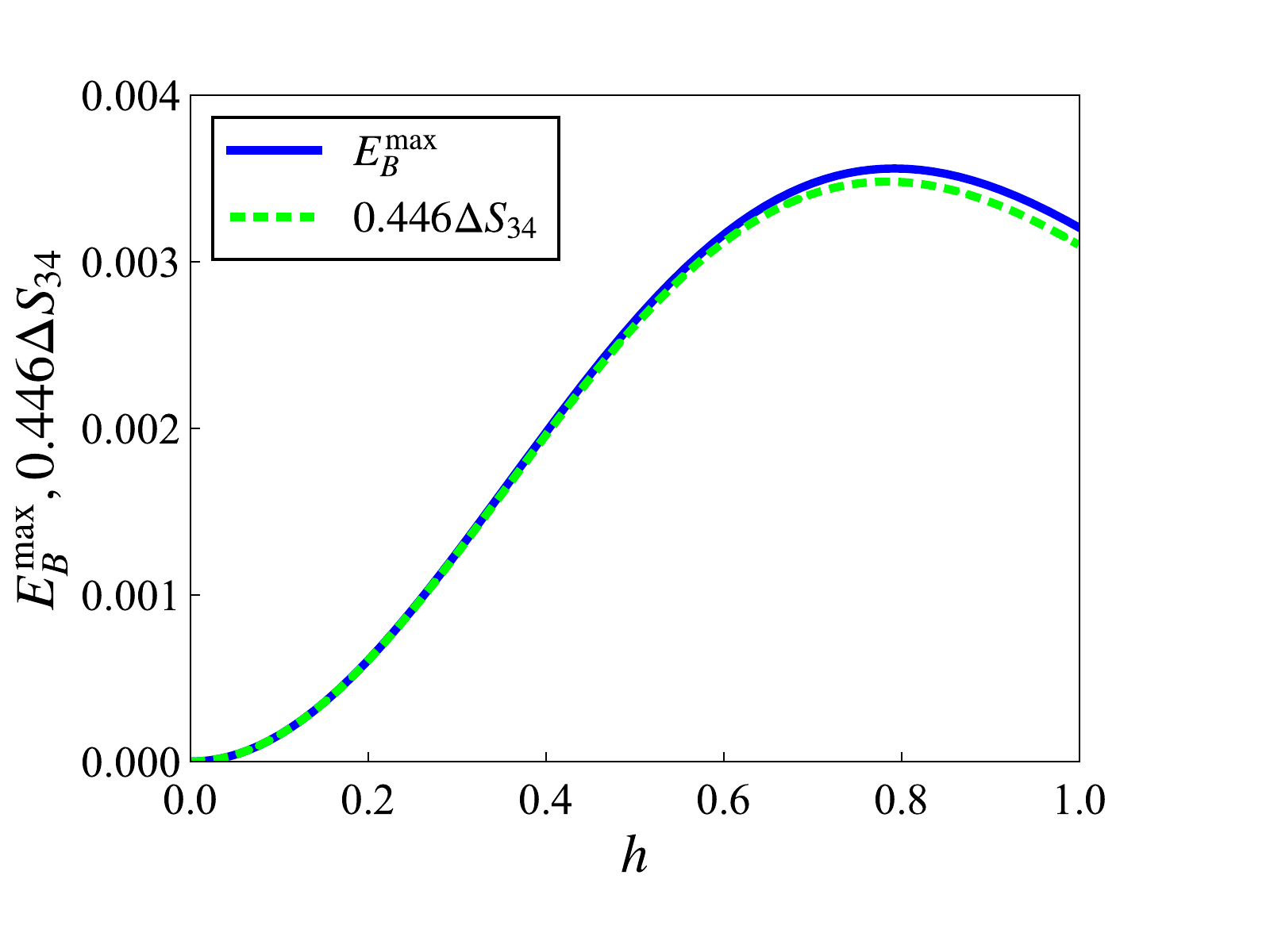}
\vspace{-0.8cm}
\caption{
$E_{B}^{\max}$ and $0.446\Delta S_{34}$ as a function of magnetic field $h$.
}
\label{EB}
\end{center}
\end{figure}

In Fig.~\ref{EB}, we present $E_{B}^{\max}$ as a function of $h$. We find that $E_B^{\max}$ increases with $h$ and takes the maximum value around $h=0.8$. 
The $h$ dependence of $E_B^{\max}$ is not similar to that of $S(\rho_{34}^{\mathrm{g}})$ shown in Fig.~\ref{S34}. 
Thus, we consider that stronger ground-state entanglement does not necessarily results in better QET performance, 
even though the entanglement is a necessary condition for realization of the QET. Instead, we find that $E_B^{\max}$ is almost proportional to the change in the von Neumann entropy of the subsystem B before and after Bob's feedback control:
\begin{eqnarray}
    E_B^{\max}\approx 0.446\times \Delta S_{34}.
    \label{EB_S34}
\end{eqnarray}
The deviation of $E_{B}^{\max}$ from $0.446\Delta S_{34}$ becomes visible for $h>0.8$, but is still less than 3.3\%. Since $I_{QC}$ is not directly related to $E_{B}^{\max}$, it is essential for energy extraction to consider the Bob's feedback control. Note that Eq.~\eqref{EB_S34} does not hold for the general $\theta$-dependent $E_B$. In other words, $E_B/\Delta S_{34}$ is generally not constant as a function of the magnetic field $h$. 
Note also that $E_{B}^{\rm{max}}$ was compared with negativity, concurrence, and quantum discord in the previous work~\cite{Trevison}, which are independent of Bob's feedback control in contrast to the entropy change $\Delta S_{34}$.

\begin{figure}[t]
\begin{center}
\includegraphics[width=8.5cm]{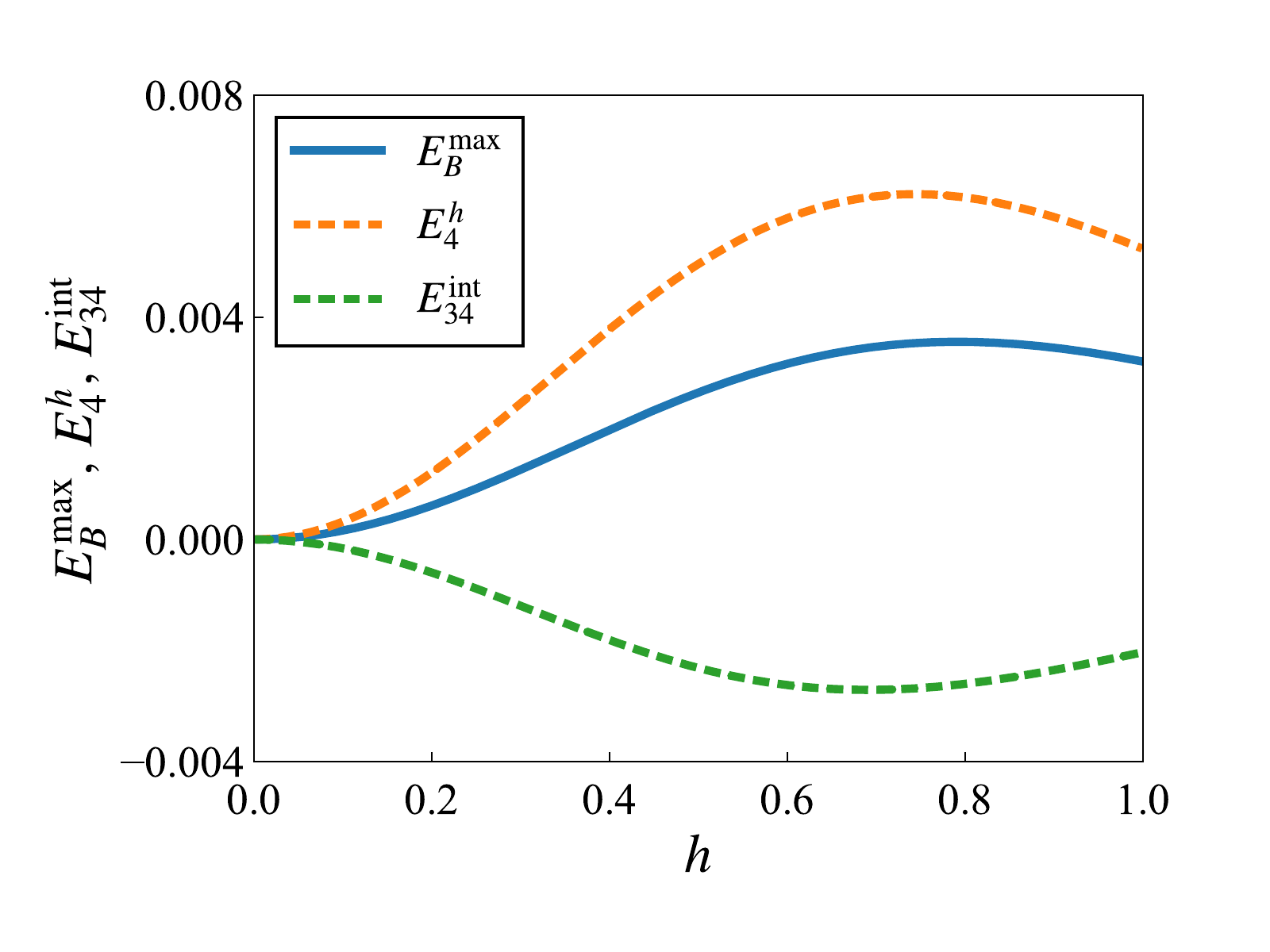}
\vspace{-0.8cm}
\caption{
$E_{B}^{\max}$, $E_4^h$ and $E_{34}^{\mathrm{int}}$ as a function of magnetic field $h$.
}
\label{EB_Eh_Eint}
\end{center}
\end{figure}

\begin{figure}[t]
\begin{center}
\includegraphics[width=8.5cm]{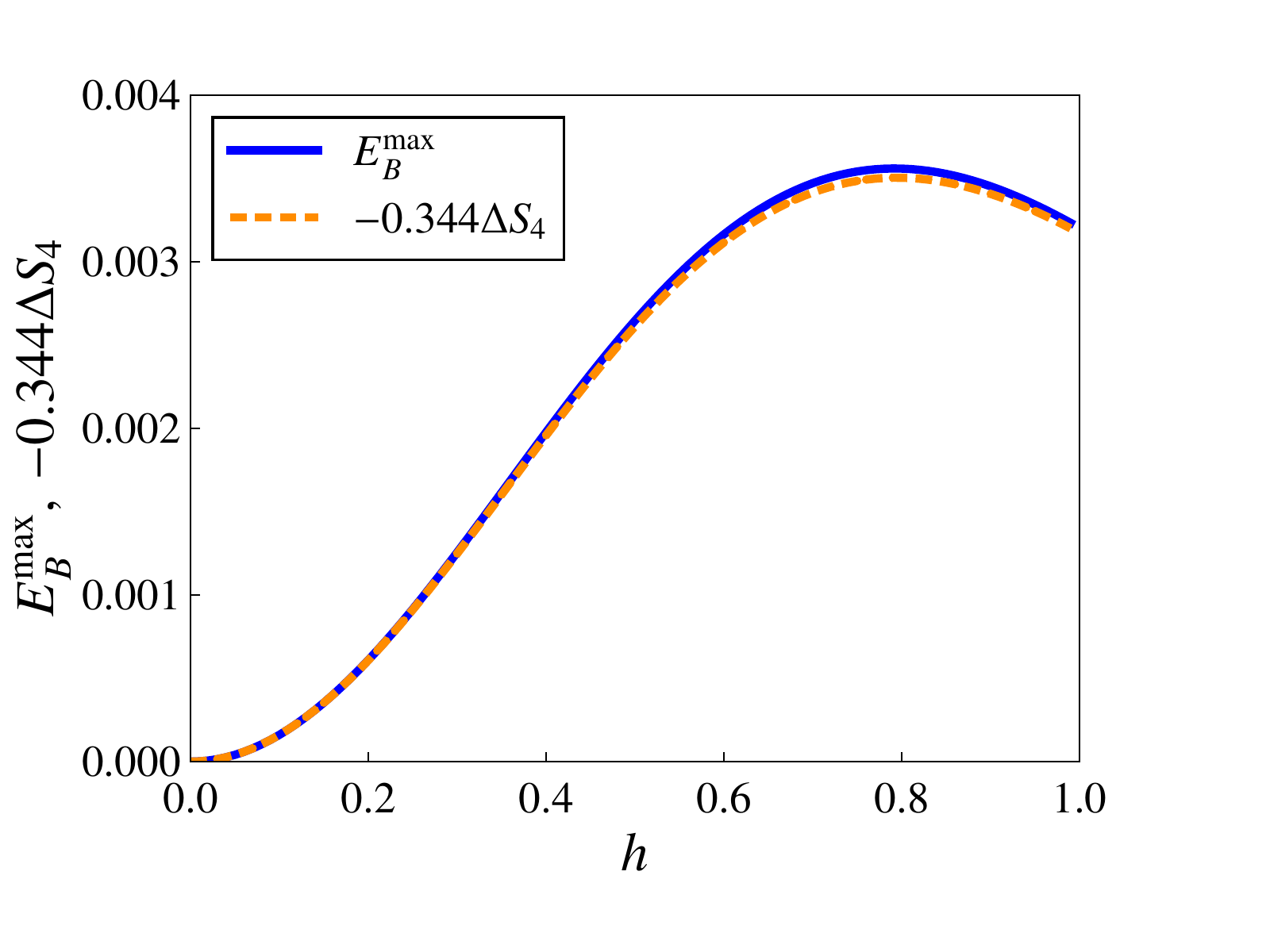}
\vspace{-0.8cm}
\caption{
$E_{B}^{\max}$ and $\Delta S_4$ as a function of magnetic field $h$.
}
\label{EB_S4}
\end{center}
\end{figure}

As already mentioned, $E_{B}^{\rm{max}}$ contains two different contributions. One is  from the magnetic-field term:
\begin{eqnarray}
E_{4}^h = - \mathrm{tr}(\rho^{\mathrm{f}}hS_4^z) 
+ \mathrm{tr}(\rho^{\mathrm{g}}hS_4^z),
\end{eqnarray}
and the other from the interaction between spins 3 and 4:
\begin{eqnarray}
E_{34}^{\mathrm{int}} = -\mathrm{tr}(\rho^{\mathrm{f}}\vec{S}_{3}\cdot\vec{S}_{4}) 
+ \mathrm{tr}(\rho^{\mathrm{g}}\vec{S}_{3}\cdot\vec{S}_{4}).
\end{eqnarray}
We separately plot their $h$ dependence in Fig.~\ref{EB_Eh_Eint}. 
We find $E_{4}^{h} \ge 0$ ($E_{34}^{\mathrm{int}} \le 0$),  which means the magnetic-field (interaction) term has the positive (negative) contribution to the energy transfer. We have numerically confirmed that $E_{34}^{\rm{int}}$ is always negative for any $\theta$ and $h$. Thus, we can never gain the extracted energy from the interaction term. Since $E_{4}^{h}$ is zero at $h=0$, we can conclude that $E_{B}^{\mathrm{max}}$ should be zero at $h=0$. Naively, QET is considered as a protocol that consumes the ground-state entanglement as a resource, and the entropy $S(\rho_{34}^{\rm{g}})$ is maximum at $h=0$. The $h$ dependence on $E_{B}^{\rm{max}}$ suggests that the effect of Bob's feedback control on $E_{B}^{\rm{max}}$ should be considered more seriously.

For comparison, we show the $h$ dependence of $\Delta S_{4}$ in Fig.~\ref{EB_S4}, where $\Delta S_{4}=S(\rho_{4}^{\rm{f}})-S(\rho_{4}^{\rm{g}})$ and the degrees of freedom at sites 1, 2, and 3 are traced out. We find that $E_{B}^{\rm{max}}$ is again proportional to $\Delta S_{4}$:
\begin{eqnarray}
    E_{B}^{\rm{max}}\approx -0.334\times\Delta S_{4},
    \label{EBmax_S4}
\end{eqnarray}
but in the present case the minus sign appears, that is, the entropy is reduced. In Eq.~(\ref{EB_S34}), we find that $\Delta S_{34}>0$ and thus the entropy production occurs by the Bob's feedback control. The sign of the entropy change depends on the position of the partition of the whole system.

\section{Discussion}
\label{sec:discussion}
Let us discuss the origin of positive $\Delta S_{34}$.
In the case of the central partition, the entropy is produced nevertheless of the Bob's feedback control. For this purpose, we investigate the nature of the effective Hamiltonians in the ground state and after Bob's operation, $H_{34}^{\rm{g}}$ and $H_{34}^{\rm{f}}$, respectively.

Bob cannot get full quantum information of Alice's side, and thus it is reasonable to consider the reduced density matrix in which the Alice's degree of freedom has been traced out. Then, the effective Hamiltonian $H_{34}^{\alpha}$ ($\alpha=\rm{g},\rm{f}$) is defined by
\begin{eqnarray}
    \rho_{34}^{\alpha}=e^{-\beta^{\alpha}H_{34}^{\alpha}}.
    \label{rho34a}
\end{eqnarray}
This can be regarded as a canonical ensemble for the effective Hamiltonian $H_{34}^{\beta}$ and the effective inverse temperature $\beta^{\alpha}$. From the definitions of $S(\rho_{34}^{\rm{g}})$, $S(\rho_{34}^{\rm{f}})$, $H_{34}^{\rm{g}}$, and $H_{34}^{\rm{f}}$, Eq.~\eqref{EB_S34} can be rewritten as
\begin{align}
    E_B^{\rm{max}}&\approx \frac{1}{\beta}(-\Delta S_{34})\label{EB_bS34}\\
    &={\rm tr}(\rho_{34}^{\rm{g}} H_{34}^{\rm{g}})-{\rm tr}(\rho_{34}^{\rm{f}} H_{34}^{\rm{f}}),
    \label{EB_Hent}
\end{align}
where we put
\begin{eqnarray}
\beta=\beta^{\rm{g}}=\beta^{\rm{f}}=-\frac{1}{0.446}.
\end{eqnarray}
Equation (\ref{EB_Hent}) has the same form as the following relation 
that holds for ordinary Hamiltonian $H_B$: $E_B^{\max}={\rm tr}(\rho_{34}^{\rm{g}} H_B)-{\rm tr}(\rho_{34}^{\rm{f}} H_B)$. The sign of $\Delta S_{34}$ naturally leads to the negative inverse temperature in Eq.~(\ref{EB_bS34}). Note that the concept of entanglement temperature is introduced in Ref.~\cite{Beny} as a proportionality coefficient between the energy cost and the extracted entropy, in which the temperature is positive. Our definition of the inverse temperature is similar to the entanglement temperature, but the magnitude and the sign of the temperature depend on the position of the partition of the whole system.

\begin{figure}[t]
\begin{center}
\includegraphics[width=8.5cm]{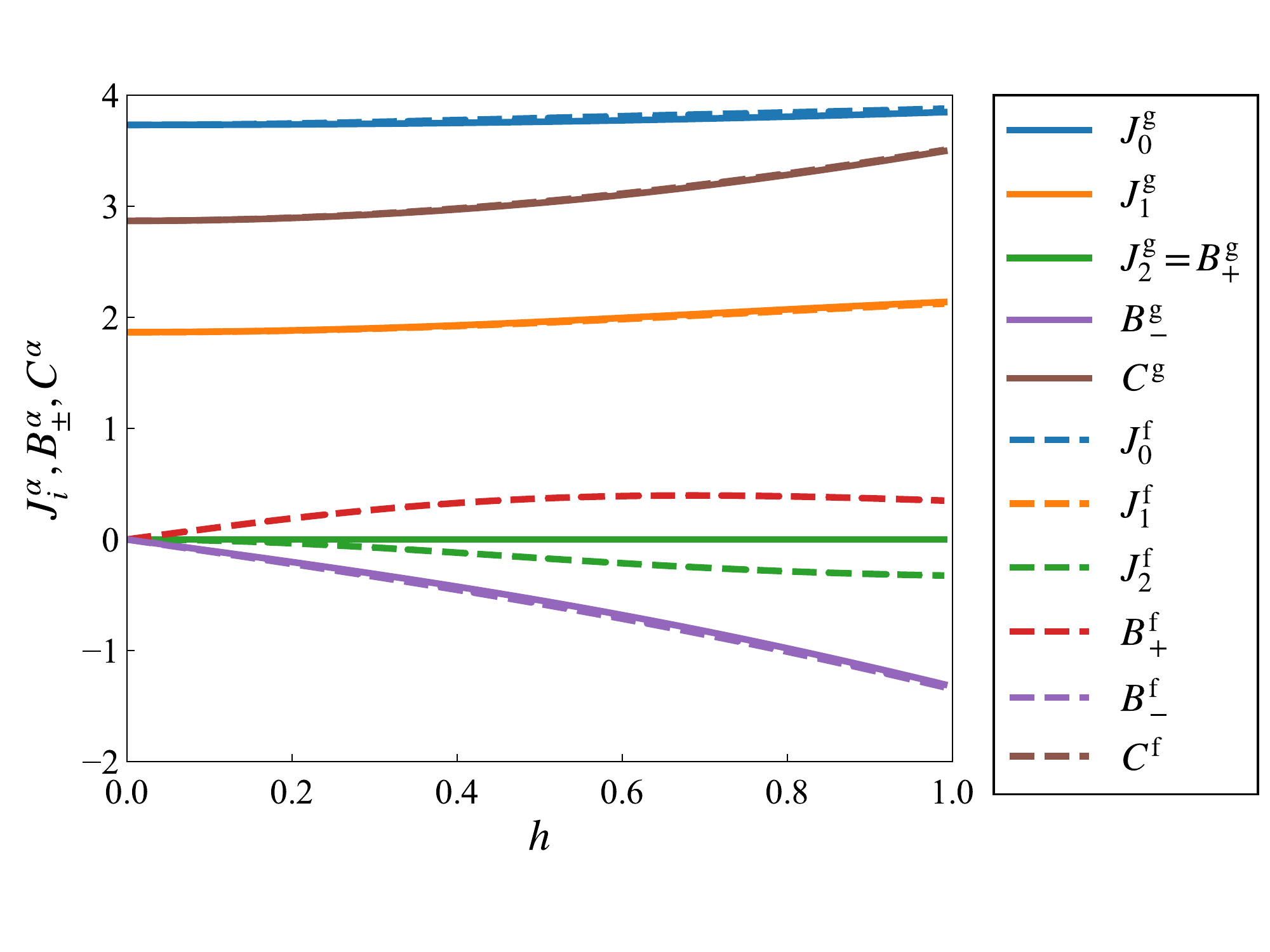}
\vspace{-0.8cm}
\caption{
 Parameters $J_{i}^{\alpha}$ ($i=0,1,2$), $B_{\pm}^{\alpha}$, and $C^{\alpha}$ in $\beta H_{34}^{\alpha}$ as a function of magnetic field $h$.
}
\label{EHalpha}
\end{center}
\end{figure}

To investigate the entropy production in detail, let us first derive the effective Hamiltonian. The operator form of the Hamiltonian is represented as
\begin{align}
\beta^{\alpha}H_{34}^{\alpha} &= J_{2}^{\alpha}(S_{3}^{+}S_{4}^{+}+S_{3}^{-}S_{4}^{-})\nonumber \\
& +J_{1}^{\alpha}(S_{3}^{+}S_{4}^{-}+S_{3}^{-}S_{4}^{+}) +J_{0}^{\alpha}S_{3}^{z}S_{4}^{z} \nonumber \\
& +B_{+}^{\alpha}(S_{3}^{z}+S_{4}^{z})+B_{-}^{\alpha}(S_{3}^{z}-S_{4}^{z})+C^{\alpha},
\label{H34eff}
\end{align}
and the Hamiltonian matrix is block-diagonalized as
\begin{align}
\beta^{\alpha}H_{34}^{\alpha}
&=\left(
\begin{matrix}\Delta_{1}^{\alpha}&0\\ 0&\Delta_{0}^{\alpha}\end{matrix}\right),\\
\Delta_{1}^{\alpha}&=\left(
\begin{matrix}
C^{\alpha}+\frac{1}{4}J_{0}^{\alpha}+B_{+}^{\alpha}&J_{2}^{\alpha}\\ J_{2}^{\alpha}&C^{\alpha}+\frac{1}{4}J_{0}^{\alpha}-B_{+}^{\alpha}
\end{matrix}
\right),\\
\Delta_{0}^{\alpha}&=\left(
\begin{matrix}
C^{\alpha}-\frac{1}{4}J_{0}^{\alpha}+B_{-}^{\alpha}&J_{1}^{\alpha}\\ J_{1}^{\alpha}&C^{\alpha}-\frac{1}{4}J_{0}^{\alpha}
-B_{-}^{\alpha}\end{matrix}\right),
\end{align}
where the basis states of $\Delta_{1}^{\alpha}$ ($\Delta_{0}^{\alpha}$) are $\left|\uparrow\uparrow\right>$ and $\left|\downarrow\downarrow\right>$ ($\left|\uparrow\downarrow\right>$ and $\left|\downarrow\uparrow\right>$). Now, we consider $\beta^{\alpha}H_{34}^{\alpha}$, not $H_{34}^{\alpha}$ itself, and then the system represented by the right hand side of Eq.~(\ref{H34eff}) behaves as an ordinary thermodynamic system with positive temperatures. 
Figure~\ref{EHalpha} shows $J_{i}^{\alpha}$ ($i=0,1,2$), $B_{\pm}^{\alpha}$, and $C^{\alpha}$ in $\beta H_{34}^{\alpha}$ as a function of magnetic field $h$. 
Note that $J_{1}^{\alpha}=J_{0}^{\alpha}/2$ for $h=0$. 
We find that $J_{2}^{\rm{f}}<0$ and $B_{+}^{\rm{f}}>0$ are induced by the Bob's operation, before which $J_{2}^{\rm{g}}=B_{+}^{\rm{g}}=0$. 
They are included only in $\Delta_{1}^{\alpha}$. 
We also find that the other parameters, $J_{1}^{\alpha}$, $J_{0}^{\alpha}$, $B_{-}^{\alpha}$, and $C^{\alpha}$ change only very little by Bob's operation although their absolute values are much larger than $J_{2}^{\rm{f}}$ and $B_{+}^{\rm{f}}$.

\begin{figure}[t]
\begin{center}
\includegraphics[width=8.5cm]{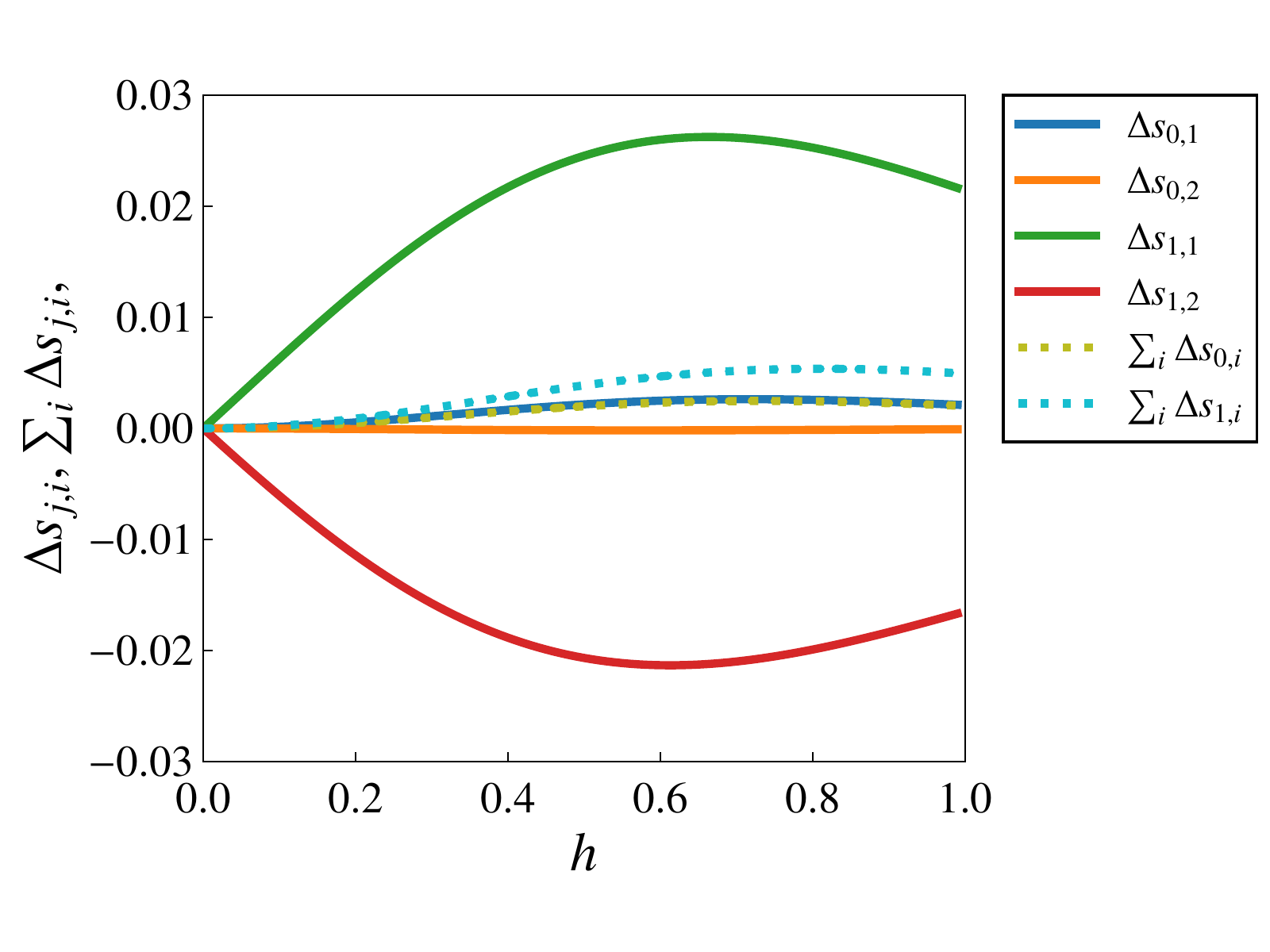}
\vspace{-0.8cm}
\caption{
 $\Delta s_{j,i (j=0,1,i=1,2)}$ as a function of magnetic field $h$.
}
\label{eigen01}
\end{center}
\end{figure}

From the eigenvalues of these effective Hamiltonians, the entropy change is evaluated as
\begin{align}
\Delta S_{34}&=\mathrm{tr}\left(\rho_{34}^{\rm{f}}\beta^{\rm{f}}H_{34}^{\rm{f}}\right)-\mathrm{tr}\left(\rho_{34}^{\rm{g}}\beta^{\rm{g}}H_{34}^{\rm{g}}\right) \nonumber \\ &=\sum_{i=1,2}\sum_{j=0,1}\Delta s_{j,i},
\label{lambda} \\
\Delta s_{j,i}&=e^{-\lambda_{j,i}^{\rm{f}}}\lambda_{j,i}^{\rm{f}}-e^{-\lambda_{j,i}^{\rm{g}}}\lambda_{j,i}^{\rm{g}}, \label{lambda2}
\end{align}
where $\lambda_{j,i}^{\alpha}$ is the $i$-th eigenvalue of $\Delta_{j}^{\alpha}$. We plot $\Delta s_{j,i}$ in Fig.~\ref{eigen01}. We find that two of the eigenvalues, $\lambda_{0,1}^{\alpha}$ (the ground state in $\beta^{\alpha}H_{34}^{\alpha}$) and $\lambda_{1,1}^{\alpha}$ (an excited state from $\Delta_{1}^{\alpha}$), positively contribute to the entropy production. 
The magnitude of $\sum_{i}\Delta s_{1,i}$ is roughly twice as large as that of  $\sum_{i}\Delta s_{0,i}$.

As a study complementary to Eq.~(\ref{lambda2}), we calculate the change in the expectation value of each term in the right hand side of Eq.~(\ref{H34eff}) before and after Bob's operation:
\begin{align}
\Delta J_{0}&=\mathrm{tr}\left(\left(\rho_{34}^{\rm{f}}J_{0}^{\rm{f}}-\rho_{34}^{\rm{g}}J_{0}^{\rm{g}}\right)S_{3}^{z}S_{4}^{z}\right), \label{eq:deltaJ0}\\
\Delta J_{1}&=\mathrm{tr}\left(\left(\rho_{34}^{\rm{f}}J_{1}^{\rm{f}}-\rho_{34}^{\rm{g}}J_{1}^{\rm{g}}\right)\left(S_{3}^{+}S_{4}^{-}+S_{3}^{-}S_{4}^{+}\right)\right) , \\
\Delta J_{2}&=\mathrm{tr}\left(\left(\rho_{34}^{\rm{f}}J_{2}^{\rm{f}}-\rho_{34}^{\rm{g}}J_{2}^{\rm{g}}\right)\left(S_{3}^{+}S_{4}^{+}+S_{3}^{-}S_{4}^{-}\right)\right) \nonumber \\
&=\mathrm{tr}\left(\rho_{34}^{\rm{f}}J_{2}^{\rm{f}}\left(S_{3}^{+}S_{4}^{+}+S_{3}^{-}S_{4}^{-}\right)\right), \\
\Delta B_{+}&=\mathrm{tr}\left(\left(\rho_{34}^{\rm{f}}B_{+}^{\rm{f}}-\rho_{34}^{\rm{g}}B_{+}^{\rm{g}}\right)\left(S_{3}^{z}+S_{4}^{z}\right)\right) \nonumber \\
&=\mathrm{tr}\left(\rho_{34}^{\rm{f}}B_{+}^{\rm{f}}\left(S_{3}^{z}+S_{4}^{z}\right)\right) , \\
\Delta B_{-}&=\mathrm{tr}\left(\left(\rho_{34}^{\rm{f}}B_{-}^{\rm{f}}-\rho_{34}^{\rm{g}}B_{-}^{\rm{g}}\right)\left(S_{3}^{z}-S_{4}^{z}\right)\right), \\
\Delta C&=C^{\rm{f}}-C^{\rm{g}}.\label{eq:deltaC}
\end{align}
By utilizing these quantities, the entropy change is represented by $\Delta S_{34}=\Delta J_{0}+\Delta J_{1}+\Delta J_{2}+\Delta B_{+}+\Delta B_{-}+\Delta C$.

\begin{figure}[t]
  \begin{minipage}[b]{1\linewidth}
    \centering
    \includegraphics[width=8.5cm]{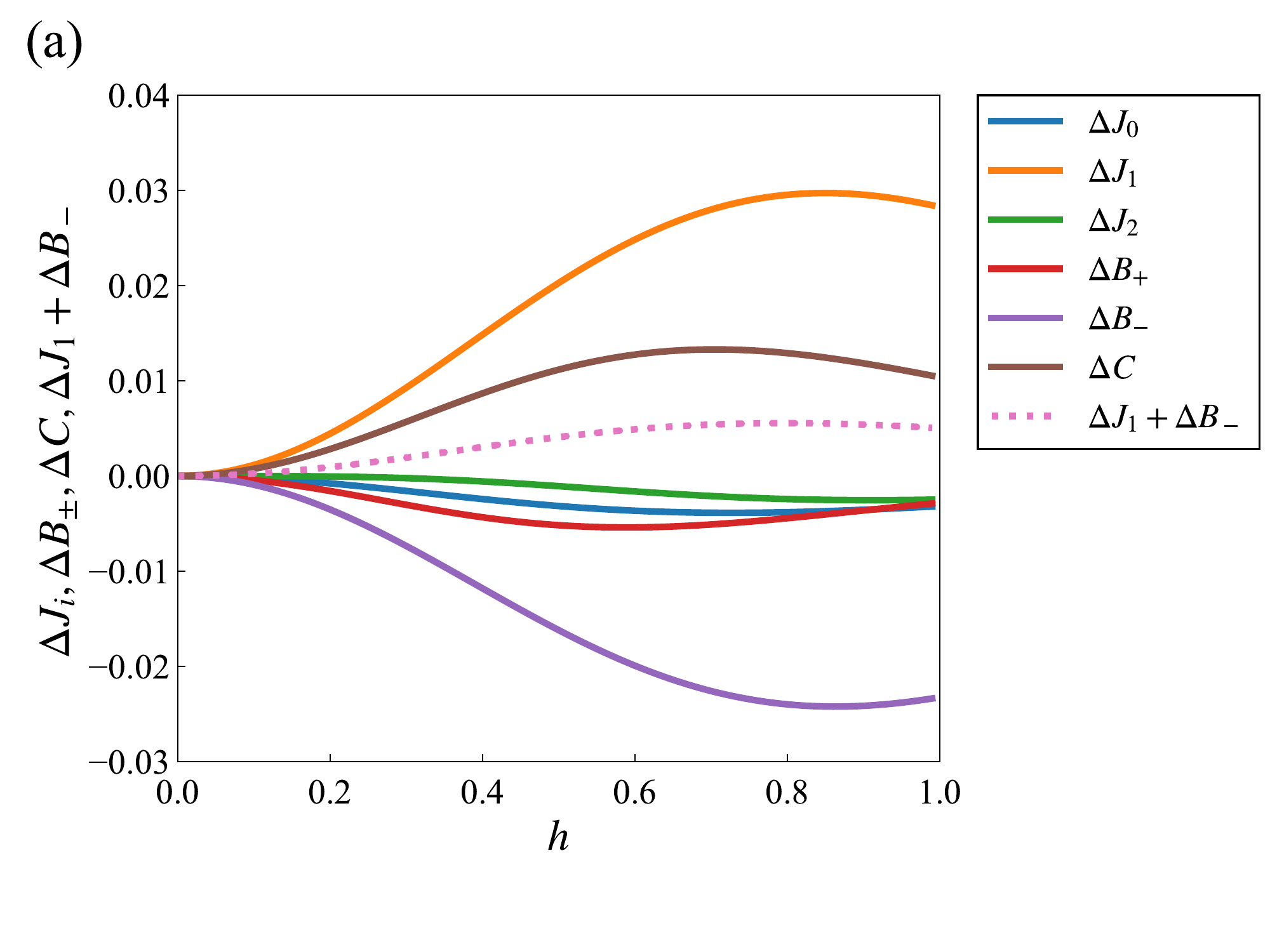}
    \vspace{-0.4cm}
  \end{minipage}\\
  \begin{minipage}[b]{1\linewidth}
    \centering
    \includegraphics[width=8.5cm]{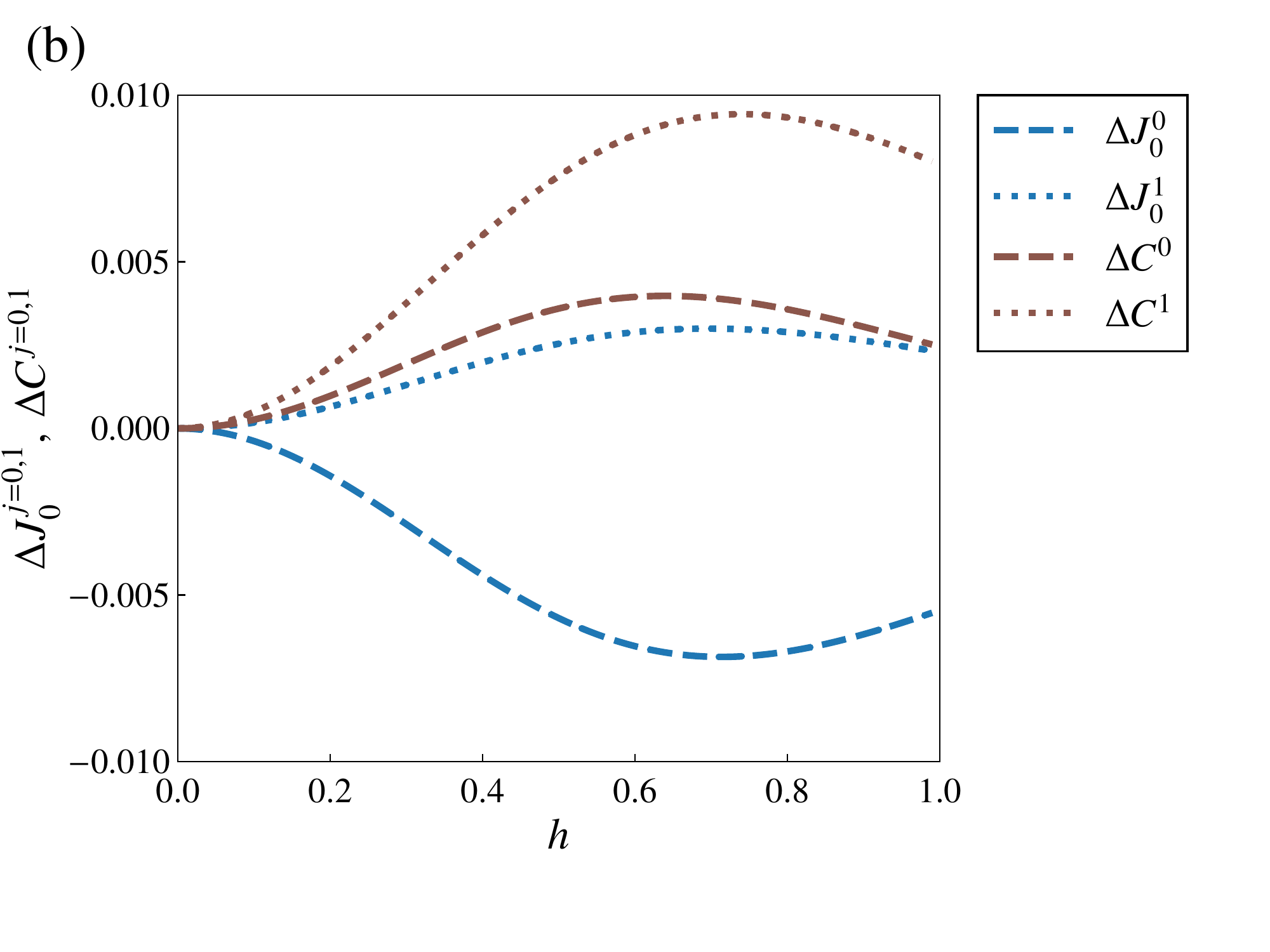}
    \vspace{-0.4cm}
  \end{minipage}
  \caption{(a) $\Delta J_{i=0,1,2}$, $\Delta B_{\pm}$, and $\Delta C$ as a function of $h$. (b) $\Delta J_{0}^{j=0,1}$ and $\Delta C^{J=0,1}$ as a function of $h$}
  \label{DH}
\end{figure}

Figure~\ref{DH}(a) shows numerical results for Eqs.~\eqref{eq:deltaJ0}--\eqref{eq:deltaC}. We find that $\Delta C$ has a positive contribution to the entropy production and shows $h$ dependence similar to that of $\Delta S_{34}$. We also find that the sum of $\Delta J_{1}$ and $\Delta B_{-}$ has a positive contribution to the entropy production too, and these quantities are determined by the eigenstates and eigenvalues from $\Delta_{0}^{\alpha}$. The quantities $\Delta J_{0}$, $\Delta J_{2}$, and $\Delta B_{+}$ are small, even though the coefficients $J_{2}^{\alpha}$ and $B_{+}^{\alpha}$ show remarkable changes by Bob's operation. The quantities $\Delta J_{2}$ and $\Delta B_{+}$ are determined by the eigenstates and the eigenvalues of $\Delta_{1}^{\alpha}$.

To investigate which matrix, $\Delta_{0}^{\alpha}$ or $\Delta_{1}^{\alpha}$, is primarily important for $\Delta C$ and $\Delta J_{0}$, we decompose these quantities into the following forms:
\begin{align}
    \Delta C&=\sum_{j=0,1}\Delta C^{j} , \\
    \Delta C^{j}&=\sum_{i=1,2}\left(C^{\rm{f}}e^{-\lambda_{j,i}^{\rm{f}}}-C^{\rm{g}}e^{-\lambda_{j,i}^{\rm{g}}}\right), \\
    \Delta J_{0}&=\sum_{j=0,1}\Delta J_{0}^{j} , \\
    \Delta J_{0}^{j}&=\sum_{i=1,2}J_{0}^{\rm{f}}e^{-\lambda_{j,i}^{\rm{f}}}\left<f,j,i|S_{3}^{z}S_{4}^{z}|f,j,i\right> \nonumber \\
    &-\sum_{i=1,2}J_{0}^{\rm{g}}e^{-\lambda_{j,i}^{\rm{g}}}\left<g,j,i|S_{3}^{z}S_{4}^{z}|g,j,i\right>,
\end{align}
where $\Delta_{j}^{\alpha}\left|\alpha,j,i\right>=\lambda_{j,i}^{\alpha}\left|\alpha,j,i\right>$. Figure~\ref{DH}(b) shows $\Delta C^{j}$ and $\Delta J_{0}^{j}$. We find that $\Delta C^{1}>\Delta C^{0}>0$. On the other hand, we find that $\Delta J_{0}^{1}>0 > \Delta J_{0}^{0}$ and finally $\Delta J_{0}$ does not contribute to the entropy production. For $\Delta C$ and $\Delta J_{0}$, the contributions to the entropy production from the excited states are larger than those from the ground state. This result is consistent with the statements below Eq.~(\ref{lambda2}).

Finally, we note that the approximate proportionality between $E_{B}^{\mathrm{max}}$ and $\Delta S_{34}$ holds when $E_B=E_B^{\rm max}$, i.e., when the QET is optimized, but not in general. 
For the derivation of Eqs.~(\ref{EB_bS34}) and (\ref{EB_Hent}), we assume that both of $\rho_{34}^{\rm{g}}$ and $\rho_{34}^{\rm{f}}$ have the same inverse temperature $\beta$. 
The same inverse temperature reminds us of isothermal processes in thermodynamics. 
However, it is unclear whether our measurement and feedback operations are quasi-static. Thus, it is an open question whether our optimized QET can be considered as an isothermal process between the thermodynamic equilibrium states for two different entanglement Hamiltonians.

\section{Summary and Future Perspectives}
\label{summary}
We have studied QET in the four-spin Heisenberg model. 
We have found that the maximum extracted energy $E_B^{\max}$ is approximately proportional to the entropy change $\Delta S_{34} ( > 0)$ and shows maximum at around $h=0.8$. The positive $\Delta S_{34}$ accounts for the entropy production in the optimized QET process. We have examined the origin of the entropy production in terms of entanglement thermodynamics based on the effective Hamiltonians. Both of the ground and excited states in the effective Hamiltonians multiplied by effective inverse temperature contribute to the entropy production, and we have found that the latter contribution is roughly twice as large as the former one. The entropy production by the latter contribution mainly originates in the constant terms in the effective Hamiltonians. Because our analysis is based on the assumption of the same effective temperature before and after the operations, it is a future work to resolve the question whether our optimized QET protocol can be viewed as an isothermal process between two equilibrium states defined by different effective Hamiltonians.

In this paper, we have considered the QET for $0<h<1$. In the case of $h>1$ where the magnitude of $S_{\rm{tot}}^{z}$ changes, a large amount of entanglement still remains in the ground state. It would be possible to utilize the entanglement for QET. It is also a future work to resolve whether the QET mechanism in this parameter region is the same as the present analysis.

Unfortunately, the efficiency for the energy transfer, given by the ratio $E_{B}^{\max}/E_{A}$, is small in the present model. The ratio is less than 1 \% at best. In this context, we need to find models that show better efficiency for the energy transfer.

The authors acknowledge S. Yamashita, T. Kumamoto, K. Ohgane, Y. Matsubayashi, T. Otaki, O. Kanehira, A. Ono, and S. Imai for fruitful discussion. This work was financially supported by JSPS KAKENHI Grant Nos. JP21K03380, JP20K03769, JP21H03455, JP21H04446, JP19K14662, JP22H01221, and CSIS, Tohoku University.

\appendix
\section{Detailed Calculations of Energies}
\label{sec:appA}
In the present Appendix, we show the details of the calculation of $E_A$ and $E_B$ for general projective measurements and feedback unitary.

First, we present calculation result of $E_A$ for general projective measurement by Alice. The measurement operator that projects spin 1 to the direction $\vec{r}$ of the Bloch sphere is written as
\begin{eqnarray}
    P_1(\mu)=\frac{1}{2}\left\{1+(-1)^\mu \vec{r}\cdot\vec{\sigma}_1\right\},
    \label{PAge}
\end{eqnarray}
where $\vec{r}=(r_x, r_y, r_z)$ is a unit real vector and $\vec{\sigma_1}=(\sigma_1^x, \sigma_1^y, 
\sigma_1^z)$. By a calculation similar to that in section I\hspace{-1.2pt}I, the infused energy by the measurement Eq. (\ref{PAge}) is given by
\begin{align}
    E_A=&\frac{1}{4}[-r_z^2\left\{ 1-4b^2-2h(c^2-d^2)-2a(c+d) \right\}\nonumber\\
    &+1-4b^2-2h(c^2-d^2)+2a(c+d)].
\end{align}

Next, we present the calculation result of the extracted energy $E_B$ for general feedback unitary by Bob.
A feedback unitary operator at site 4 is generally written as
\begin{eqnarray}
U_4(\mu)=\cos{\omega_\mu}+i\vec{n}_\mu\cdot\vec{\sigma}_4 \sin{\omega_\mu},
\end{eqnarray}
and the extracted energy is evaluated as
\begin{align}
E_B&=\sum_{\mu}\left(x(\mu)\cos{2\omega_\mu}+y(\mu)\sin{2\omega_\mu}+z(\mu)\right)\nonumber\\
    &\le\frac{1}{16}\sum_{\mu}\left(\sqrt{x(\mu)^2+y(\mu)^2}+z(\mu)\right),
\end{align}
where $x(\mu)$, $y(\mu)$, and $z(\mu)$ are given by
\begin{align}
    x(\mu)&=(-1)^\mu[2r_z\{n_{z\mu}^2(-4ha^2-4hb^2\nonumber\\&+(2h-1)c^2+(2h+1)d^2+2a(c-d))\nonumber\\
    &+4ha^2+4hb^2-(2h-1)c^2-(2h+1)d^2\nonumber\\
    &+2a(c-d)\}\nonumber\\
    &-4(r_xn_{x\mu}-r_yn_{y\mu})n_{z\mu}b(c-d)]\nonumber\\
    &-2n_{z\mu}^2\{1-4b^2-2h(c^2-d^2)-2a(c+d)\}\nonumber\\
    &+1-4b^2-2h(c^2-d^2)+4\eta,\\
    y(\mu)&=(-1)^\mu4b(r_xn_{y\mu}-r_y n_{x\mu})(4ha+c-d),\\
    z(\mu)&=(-1)^\mu[2r_z\{n_{z\mu}^2(4ha^2+4hb^2\nonumber\\&-(2h-1)c^2-(2h+1)d^2-2a(c-d))-2a(c+d)\nonumber\\
    &+4\eta(c^2-d^2)\}\nonumber\\
    &+4(r_x n_{x\mu}-r_y n_{y\mu})n_{z\mu}b(c-d)]\nonumber\\
    &+2n_{z\mu}^2\{1-4b^2-2h(c^2-d^2)-2a(c+d)\}\nonumber\\
    &-1+4b^2+2h(c^2-d^2)-4\eta,
\end{align}
and equality holds at $\omega_\mu=\frac{1}{2}\arctan{\frac{Y(\mu)}{X(\mu)}}$. Extracted energy $E_B$ is maximal when $\vec{r}=(\cos{t},\sin{t},0)$, $\vec{n}_\mu=((-1)^{N_\mu}\sin{t},(-1)^{N_\mu+1}\cos{t},0)$, with real parameter $t$ and $N_\mu=0,1$.

\section{Minimal QET Model Revisited}
\label{sec:appB}

We would like to mention that
Eq.~(\ref{EB_bS34}) is also satisfied in the minimal QET model. The Hamiltonian of the QET is defined by
\begin{eqnarray}
H=h\sigma_{A}^{z}+h\sigma_{B}^{z}+2k\sigma_{A}^{x}\sigma_{B}^{x}+{\rm const.},
\end{eqnarray}
where the constant term is determined by Eq.~(\ref{geta}). The extracted energy after the Bob's operation is obtained as
\begin{eqnarray}
E_{B}^{\max}=\frac{h^{2}+2k^{2}}{\sqrt{h^{2}+k^{2}}}\left[\sqrt{1+\frac{h^{2}k^{2}}{(h^{2}+2k^{2})^{2}}}-1\right].
\end{eqnarray}
To evaluate the entropy, the reduced density matrices for the ground state and the state after the Bob's operation, in which the degree of site A has been traced out, are respectively given by
\begin{eqnarray}
    \rho_{B}^{\rm{g}}=\left(\begin{matrix}
        a^{2}&0\\ 0&b^{2}
    \end{matrix}\right),
    \label{minrhog}
\end{eqnarray}
and
\begin{eqnarray}
    \rho_{B}^{\rm{f}}=\left(\begin{matrix}
        a^{2}\cos^{2}\theta+b^{2}\sin^{2}\theta&0\\ 0&b^{2}\cos^{2}\theta+a^{2}\sin^{2}\theta
    \end{matrix}\right),
    \label{minrhof}
\end{eqnarray}
where their basis states are $\left|\uparrow\right>$ and $\left|\downarrow\right>$, and the parameters, $a$, $b$, and $\theta$, are defined by
\begin{align}
    a&=\frac{1}{\sqrt{2}}\sqrt{1-\frac{h}{\sqrt{h^{2}+k^{2}}}}, \\
    b&=\frac{1}{\sqrt{2}}\sqrt{1+\frac{h}{\sqrt{h^{2}+k^{2}}}}, \\
    \cos(2\theta)&=\frac{h^{2}+k^{2}}{\sqrt{\left(h^{2}+2k^{2}\right)^{2}+h^{2}k^{2}}}.
\end{align}
Here, the angle $\theta$ is introduced in the unitary operation $U_{B}(\mu)=\cos\theta + i(-1)^{\mu}\sigma_{B}^{y}\sin\theta$. Then, the entropy is calculated as $S(\rho_{B}^{\alpha})=-\mathrm{tr}_{B}\left(\rho_{B}^{\alpha}\log\rho_{B}^{\alpha}\right)$, and we obtain $\Delta S_{B}=S(\rho_{B}^{\rm{f}})-S(\rho_{B}^{\rm{g}})$. Since the density matrices have been already diagonalized, the derivation of the effective Hamiltonian is straightforward.

\begin{figure}[h]
\begin{center}
\includegraphics[width=8.5cm]{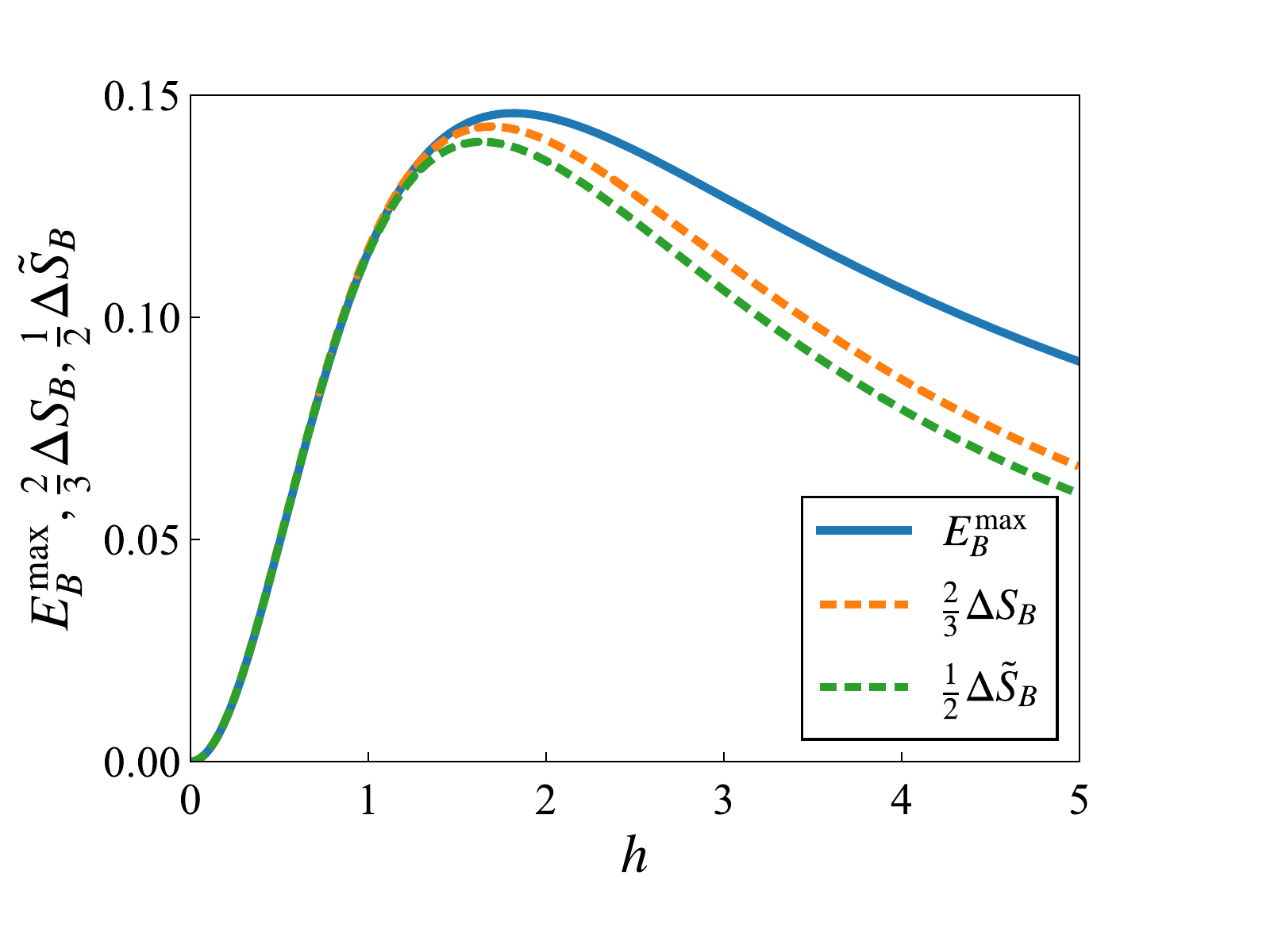}
\vspace{-0.8cm}
\caption{
$E_{B}^{\rm{max}}$, $\frac{2}{3}\Delta S_B$ and $\frac{1}{2}\Delta\tilde{S}_B$ as a function of magnetic field $h$ for $k=1$. 
}
\label{EB_minimalQET}
\end{center}
\end{figure}

Figure~\ref{EB_minimalQET} shows the numerical results. We find
\begin{eqnarray}
    E_B^{\max}\approx -\frac{\Delta S_{B}}{\beta}
    =-\frac{S_{B}^{\rm{f}}-S_{B}^{\rm{g}}}{\beta},
    \label{EB_SB}
\end{eqnarray}
where $0\le h<2$ and $\beta\approx -\frac{3}{2}$. The result is consistent with Eqs.~(\ref{EB_bS34}) and (\ref{EB_Hent}), and the effective inverse temperature $\beta$ is again negative. In the minimal QET case, there is no off-diagonal component of $\rho_{B}^{\alpha}$. By looking at $\rho_{B}^{\alpha}$ in Eqs.~(\ref{minrhog}) and (\ref{minrhof}), we clearly observe that the two diagonal components become closer after the rotation of spin B due to the Bob's operation $U_{B}(\mu)$. This is a source of the entropy production. Precisely speaking, the mechanism of the entropy production in the minimal QET should be distinguished from that in the four-spin QET in which the behavior of the spin next to the Bob's spin determines the entropy production, although in both cases the Bob's operation is crucial.

We also found that
\begin{eqnarray}
     E_B^{\max}
     \approx -\frac{\Delta\tilde{S}_B}{\tilde{\beta}}
     =-\frac{\tilde{S}_{B}^{\rm{f}}-S_{B}^{\rm{g}}}{\tilde{\beta}},
     \label{EB_SBt}
\end{eqnarray}
where
\begin{eqnarray}
    \tilde{S}_{B}^{\rm{f}}=-\rm{tr}(\rho_{B}^{\rm{f}}\log\rho_{B}^{\rm{g}}),
\end{eqnarray}
and $\tilde{\beta}\approx -2$. In this case, $\tilde{\beta}$ is also negative. Figure~\ref{EB_minimalQET} shows the comparison between $E_{B}^{\max}$ , $\frac{2}{3}\Delta S_{B}$ and $\frac{1}{2}\Delta\tilde{S}_B$.

\end{document}